\documentclass[a4paper, amsfonts, amssymb, amsmath, reprint, showkeys, nofootinbib, twoside]{revtex4-1}
\usepackage[english]{babel}
\usepackage[utf8]{inputenc}
\usepackage[colorinlistoftodos, color=green!40, prependcaption]{todonotes}
\usepackage{amsthm}
\usepackage{mathtools}
\usepackage{physics}
\usepackage{xcolor}
\usepackage{graphicx}
\usepackage[left=20mm,right=20mm,top=25mm,columnsep=15pt]{geometry} 
\usepackage{adjustbox}
\usepackage{placeins}
\usepackage[T1]{fontenc}
\usepackage{lipsum}
\usepackage{csquotes}
\usepackage{comment}
\usepackage{enumitem}
\usepackage[pdftex, pdftitle={Article}, pdfauthor={Author}]{hyperref} 

\newcommand{\tg}[1]{\textcolor{black}{#1}}

\begin{document} 

\title{Acoustic modulation of shear thickening transition in dense adhesive suspensions}

\author{Aoxuan Wang$^{1,2}$, Fabrice Toussaint$^2$, Thomas Gibaud$^{1,3}$}
    \email[Corresponding author, ]{thomas.gibaud@ens-lyon.fr}
    \affiliation{$^1$ENSL, CNRS, Laboratoire de Physique, F-69342 Lyon, France}
    \affiliation{$^2$Holcim Innovation Center, Lyon, France}
    \affiliation{$^3$Department of Polymer Engineering, IPC, University of Minho, Guimarães, 4804-533 Portugal}
\date{\today} 

\begin{abstract}
\tg{Discontinuous shear thickening (DST) in dense suspensions leads to flow instabilities that limit processing in many systems. While high-power ultrasound has been reported to reduce the apparent viscosity of such materials, the origin of this effect remains unclear. Here, we investigate dense adhesive cornstarch suspensions, where shear thickening arises from fragile, load-bearing force networks embedded in heterogeneous density-wave structures.
Using a rheo-ultrasound setup, we show that ultrasound does not directly reduce viscosity but instead shifts the shear-thickening transition toward higher shear rates. This is evidenced by the collapse of stress probability distributions onto master curves, revealing a continuous evolution toward more fluid-like states without a sharp threshold.
We interpret these results through a separation of time scales, in which the suspension behaves as an effectively immobile porous medium subjected to high-frequency interstitial flows. Fluidization then arises from a combination of boundary slip, bulk destabilization of force networks by drag-force fluctuations, and localized acoustic streaming.
Beyond these mechanisms, we propose that ultrasound modifies the stability of force networks by introducing fluctuating hydrodynamic forces at the pore scale. As a result, larger stresses or shear rates are required to sustain jammed states, leading to a continuous renormalization of the DST transition. These findings provide a consistent physical picture of acoustic fluidization in adhesive suspensions and establish ultrasound as a powerful tool to control discontinuous shear thickening in confined flows.}

\end{abstract}

\maketitle

\section{Introduction}

\tg{High-power ultrasound has emerged as a versatile tool to fluidize and restructure soft matter systems. In colloidal gels, composed of sticky sphere at low volume fractions, acoustic forcing promotes softening through the nucleation and propagation of microcracks within the load-bearing network~\cite{gibaud2020, dages2021}. In dense suspensions of non-Brownian rough 
particles, shear thickening -- a substantial increase in viscosity under high shear forces~\cite{Barnes1989, Wagner2009, Mari2014} -- is driven by a stress-induced transition from lubricated to frictional contacts: below a critical stress, particles are separated by a hydrodynamic film and flow easily, whereas above it, surface asperities interlock, frictional contacts proliferate, and the viscosity rises sharply~\cite{wyart2014, Mari2014}. When this transition is discontinuous (DST), 
the suspension can jam abruptly, causing catastrophic failure of pumping and mixing equipment~\cite{Gurgen2017}. In the hard-sphere limit, ultrasound has been shown to reduce viscosity when the vibration amplitude exceeds the particle surface roughness~\cite{sehgal2019}, although the precise mechanism remains unclear: ultrasound must generate not only large stresses but also sufficient force gradients along the chain — if the acoustic forcing is spatially uniform on the scale of the 
chain, the entire force network oscillates nearly as a rigid body and remains intact; only spatial inhomogeneities in the acoustic forcing can create the internal deformation needed to separate particle contacts.}

\tg{These two cases — colloidal gels and dense shear thickening frictional hard-sphere suspensions — represent opposite limits of interparticle interaction: strongly attractive on one end, purely repulsive on the other. An intermediate and practically important class of systems lies between these limits: dense suspensions in which short-range attractions are too weak to drive gelation at rest, yet strong enough to participate in shear thickening under flow. In such systems, adhesive contacts between particles reinforce frictional force chains, giving rise to a richer and more heterogeneous DST phenomenology characterized by large stress fluctuations, density waves, and intermittent transitions between fluid-like and jammed states~\cite{Lootens2003, saint2018, gauthier2023}. Cornstarch suspensions are a canonical example of this intermediate case, where polysaccharide chains coating the granule surfaces entangle and form hydrogen-bonded contacts under compression or shear, reinforcing frictional force chains with adhesive bonds that make jammed configurations mechanically persistent~\cite{brown2014, oyarte2017, gauthier2023}. Recent studies have demonstrated acoustic control of viscosity in dense suspensions, including hard-sphere colloidal systems~\cite{sehgal2019, sehgal2024} and, at high concentrations in the shear-jamming regime, cornstarch suspensions~\cite{ong2024jamming}. However, the effect of ultrasound on cornstarch in the DST regime — where the suspension exhibits density-wave dynamics~\cite{gauthier2023} and intermittent stress fluctuations at concentrations well below shear jamming, and where adhesive interactions fundamentally alter the force-chain dynamics — remains unexplored.}


\tg{The DST phenomenology of adhesive dense suspensions is directly relevant to cementitious slurries — dense suspensions of mildly adhesive particles that remain flowable prior to solidification yet exhibit pronounced shear thickening that complicates mixing, pumping, and extrusion operations~\cite{feys2008,feys2009}. In such materials, the ability to tune viscosity in situ without modifying the suspension chemistry or interrupting the process is of considerable practical value. 
Here we address this question directly, using cornstarch suspensions as a model system and a custom rheo-ultrasound platform to probe how acoustic forcing alters the onset and dynamics of DST under controlled shear. We quantify how ultrasound shifts the critical shear rate and reshapes the intermittent stress response, and propose a physical scenario in which ultrasound-induced interstitial flows generate force gradients that continuously destabilize load-bearing structures, tuning the transition between fluid-like and jammed states.}

\section{Results}

\begin{figure}[htbp]
    \centering
    \includegraphics[width=0.47\textwidth]{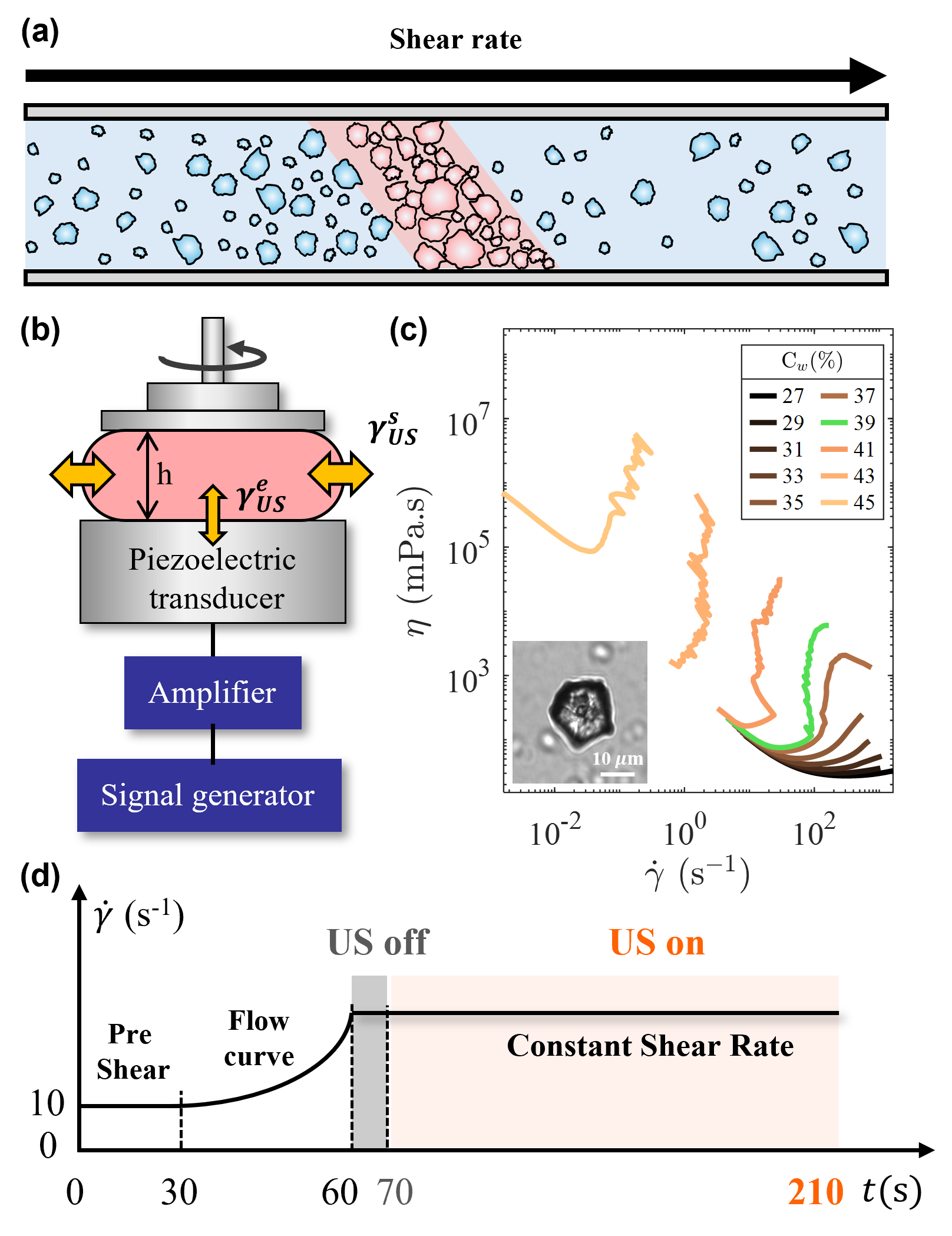}
    \caption{
        (a) Shear thickening model for corn starch suspensions.
        (b) Rheo-US apparatus.
        (c) Flow curve ($\eta$ vs $\dot\gamma$) for cornstarch suspensions at $c_w=$ 27-45 \%. Inset : bright field microscopy image of a corn starch particle.        
        (d) Rheology protocol based on imposed $\dot{\gamma}$.
    }
    \label{fig:1fc}
\end{figure}

\subsection{Rheo-US: integrating ultrasound with shear rheology in dense suspensions. }
We work with cornstarch particles (Sigma-Aldrich) dispersed in water and density-matched using cesium chloride, following the protocol in~\cite{han2016,saint2018}. Using bright-field microscopy, we determine that the particles are polydisperse with an average radius $r_{cs} = 7~\mu\mathrm{m}$. Surface roughness (rugosity) is quantified as the standard deviation $\delta_h$ of the contour length relative to a circular particle of radius equal to the mean contour length. We find that the roughness is proportional to the particle size, $\delta_h = 0.09r_{cs}$, yielding an average value of $\delta_h = 0.56~\mu\mathrm{m}$ (see SM.A and B~\cite{SM_wang}).
Suspensions are loaded into the Rheo-US apparatus (schematic in Fig.~\ref{fig:1fc}(b)), enabling rheological measurements under ultrasonic vibrations~\cite{gibaud2020,dages2021}. The apparatus consists of a stress-controlled rheometer (Anton Paar MCR 301) with a plate geometry (radius $R = 25$~mm, gap $h = 1$~mm). The bottom plate is a piezoelectric transducer (ThorLab PKT40B, radius 30~mm) driven by an amplified sinusoidal voltage $U$, converting electrical input into surface mechanical vibrations of amplitude $a_{US}$ measured with a laser vibrometer (Polytec OVF-505). we work at the resonance frequency of the transducer, $f = 15.17$~kHz. $a_{US}$ is spatially uniform except at the center ($\simeq1.5$ increase).
Calibration of $a_{\mathrm{US}}$ with input voltage $U$ shows that we can reach amplitudes up to $4~\mu\mathrm{m}$. Within this range, heating remains minimal:  $+2~^\circ\mathrm{C}$ over 150~s—the duration of our experiments. $a_{\mathrm{US}}$ remains stable over 150~s, except at the highest values, where a $5\%$ decrease is observed (see SM.C~\cite{SM_wang}).

The flow curves of cornstarch suspensions with no-Ultrasound (US) at various weight concentrations, ranging from $c_w = 27\%$ to $45\%$, are shown in Fig.~\ref{fig:1fc}(c). At low shear rates $\dot{\gamma}$, all suspensions exhibit shear thinning: the viscosity $\eta$ decreases with increasing $\dot{\gamma}$. As discussed in~\cite{gauthier2023}, this behavior arises from adhesive forces between particles, estimated as $F_{\mathrm{adh}} \approx 20~\mathrm{nN}$. At higher shear rates, the suspensions exhibit shear thickening. Above $c_w = 36.5\%$, we observe a transition to discontinuous shear thickening (DST). In the following, we focus on the suspension at $c_w = 39\%$.


\begin{figure*}[htbp]
    \centering
    \includegraphics[width=1\textwidth]{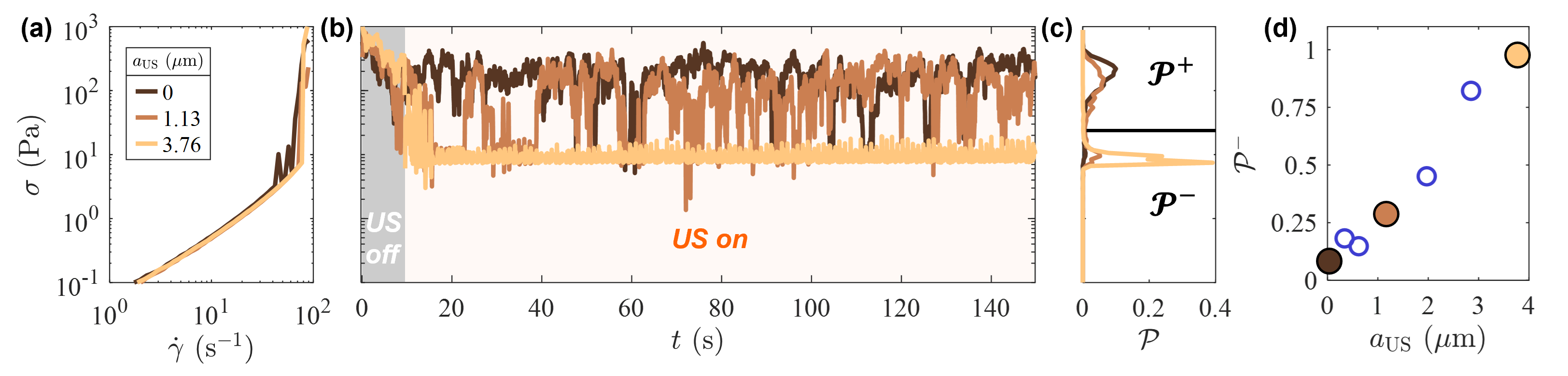}
    \caption{ Effect of high power ultrasound on the giant stress fluctuations. 
        (a) Superposition of three flow curves obtained by following the protocol described in Fig.~\ref{fig:1fc}, for a cornstarch suspensions at $c_w=39$~\%w. 
        (b) Shear stress response under constant shear rate $\dot\gamma = 90$ ~s$^{-1}$ for three ultrasound amplitudes: $a_\mathrm{US} = 0$ (black), $1.23$ (brown), and $3.76$ $\mu$m (orange). The stress is recorded at 20~Hz. 
        (c) Probability distribution function $\mathcal{P(\sigma)}$ of shear stress, computed from the stress fluctuations measured during the constant shear rate phase (from $t = 10$~s to $t = 150$~s). The distribution is binarized using a stress threshold $\sigma_t = 28$~Pa and integrated such that $\mathcal{P}^+$ + $\mathcal{P}^-$ = 1. Here, $\mathcal{P}^-$ is the probability of observing a stress below $\sigma_\mathrm{t}$, and $\mathcal{P}^+$ is its complementary probability.
        (d)  Evolution of $\mathcal{P}^-$ with ultrasound amplitude $a_\mathrm{US}$ at a shear rate of $\dot\gamma = 90$~s$^{-1}$. The colored circles correspond to the experiments in (c). The empty circles are additional measurements.
    }
    \label{fig:2probablity}
\end{figure*}
\tg{To investigate the effect of ultrasound on shear-thickening behavior, we impose a constant shear rate rather than a constant stress, despite DST being intrinsically a stress-driven phenomenon~\cite{wyart2014, Mari2014}. This choice is motivated by three reasons.
First, the DST flow curve has a region of negative slope ($d\sigma/d\dot\gamma < 0$), which is mechanically unstable under stress control: the system jumps discontinuously between the low- and high-viscosity branches, suppressing the intermittent stress fluctuations that are the central observable of this study. 
Imposing the shear rate instead indeed allows the system to reside in this unstable region and exhibit viscosity bifurcationy as shown in~\cite{Lootens2003}, from which the probability distributions $P^{\pm}$  can be meaningfully computed over the duration of each experiment. 
Second, since ultrasound imposes a kinematic perturbation—a surface displacement amplitude $a_\mathrm{US}$ corresponding to an oscillatory strain—controlling the shear rate keeps both perturbations in the same kinematic space, allowing us to interpret their combined effect through a single shifted critical shear rate $\dot\gamma^*_\mathrm{US}$. Mixing stress control (rheometer) with strain control (ultrasound) would introduce two conjugate variable pairs simultaneously, making it difficult to disentangle changes in the mechanical response unambiguously to either perturbation. Third, under rate control, both the shear stress $\sigma$ and the normal force $F_N$ are free to fluctuate simultaneously, providing two independent diagnostics of the jammed state (see SM.D~\cite{SM_wang}). Fourth most industrail process such as pumping or mixing are shear rate controlled.}

Accordingly, as shown in Fig.~\ref{fig:1fc}(d), the protocol begins with the application of a low constant shear rate to homogenize the suspension. We then perform a flow curve by logarithmically increasing the shear rate from $1$ to the target value, which is subsequently held constant. In this final stage, ultrasound is applied for $140$~s at constant amplitude $a_{\mathrm{US}}$, starting at $t = 10$~s. Throughout the protocol, we record the resulting shear stress $\sigma$ and normal force $F_N$. In the final stage, the acquisition frequency is set to 20~Hz. The results of this protocol are shown in Fig.~\ref{fig:2probablity} for different ultrasonic amplitudes: $a_\mathrm{US} = 0$, $1.13$, and $3.76~\mu\mathrm{m}$, at a constant shear rate of $\dot{\gamma} = 90~\mathrm{s}^{-1}$. As shown in the flow curves of Fig.~\ref{fig:2probablity}(a), the shear rate at which the system enters the DST regime is not identical across runs. We thereafter define a critical shear rate $\dot\gamma^{*}$, corresponding to the onset of the high-stress regime, and use it to normalize the shear rate in subsequent analyses (see SM.D~\cite{SM_wang} for the variations $\dot\gamma^{*}$ across different runs).

{\subsection{Experimental evidence of acoustic modulation of shear thickening transition.}

\tg{In the absence of ultrasound ($a_\mathrm{US} = 0$), the system exhibits large temporal fluctuations of both shear stress (Fig.~\ref{fig:2probablity}(b)) and normal force (see SM.D~\cite{SM_wang}). Such behavior, previously reported in dense suspensions~\cite{Lootens2003,gauthier2023}, is characteristic of intermittent dynamics, i.e. irregular switching in time between distinct mechanical states, associated with the formation of heterogeneous density or stress waves and resulting in strong spatiotemporal fluctuations.
These fluctuations are quantified through the stress probability distribution function $\mathcal{P}$ (Fig.~\ref{fig:2probablity}(c)).  $\mathcal{P}$ is bimodal: the system alternates between two states separated by a threshold $\sigma_t=28$~Pa: a low-stress, high-fluidity state with probability $\mathcal{P}^-$, and a high-stress, low-fluidity state with probability $\mathcal{P}^+$, such that $\mathcal{P}^- + \mathcal{P}^+ = 1$. In the absence of ultrasound, the distribution is dominated by the high-stress state, i.e. $\mathcal{P}^+$ is maximal while $\mathcal{P}^-$ is minimal, reflecting the predominance of jammed configurations.
As $a_\mathrm{US}$ increases (from 0 to $3.76~\mu$m), there is a  a redistribution toward $\mathcal{P}^+$ as the system progressively shifts toward the low-stress state. Although intermediate amplitudes (e.g., $a_\mathrm{US}=1.13~\mu$m) may display a more balanced or visibly bimodal distribution, this corresponds to a redistribution between the two states rather than an increase in fluctuation amplitude. This evolution is captured quantitatively by the monotonic increase of $\mathcal{P}^-$ with $a_\mathrm{US}$ (Fig.~\ref{fig:2probablity}(d)), demonstrating that ultrasound progressively suppresses intermittent high-stress events and promotes flow.}


 \begin{figure}[htbp]
    \centering
    \includegraphics[width=0.47\textwidth]{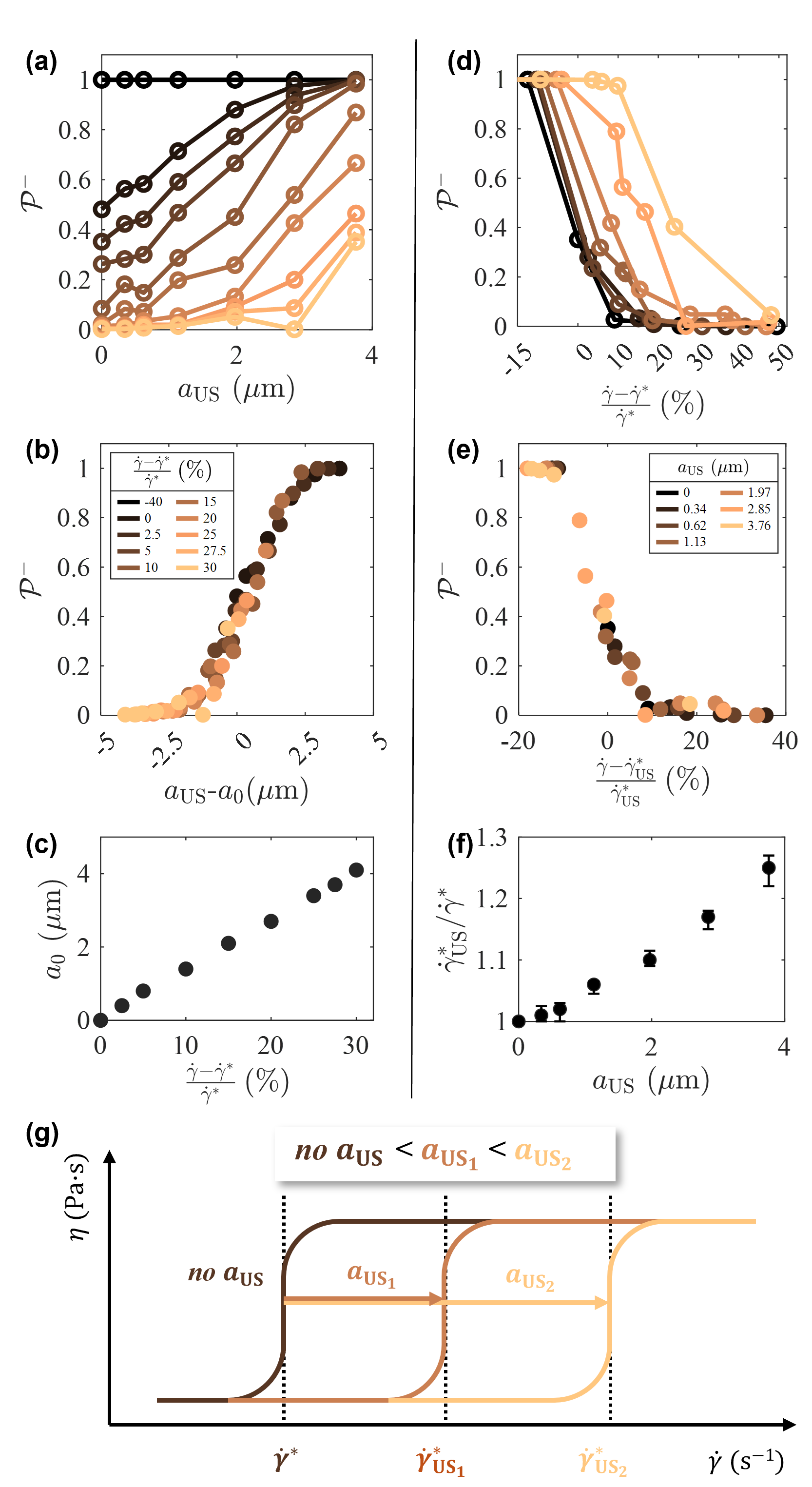}
    \caption{
        Influence of ultrasound on the DST.  Left side (a-c): constant $\dot\gamma$ series.  Right side (d-e): constant $a_{US}$ series.
        (a) Probability $\mathcal{P}^-$ of the low-shear-stress state versus ultrasound amplitude $a_\mathrm{US}$ at constant shear rates $\dot{\gamma}$ normalized by the critical shear rate $\dot\gamma^{*}$ (where $\sigma(\dot\gamma^{*})=28$~Pa, see Fig.~\ref{fig:2probablity}(a)).  
        (b) Scaling of $\mathcal{P}^-$ with $a_\mathrm{US} - a_0$, where $a_0$ is a shift parameter with respect to $\mathcal{P}^-(\dot\gamma^{*}=0)$. 
        (c) Variation of $a_0$ with the normalized shear rate with respect to $\dot\gamma^{*}$.
        (d) $\mathcal{P}^-$ versus normalized shear with respect to $\dot\gamma^{*}$ at fixed $a_\mathrm{US}$.  
        (e) Scaling of $\mathcal{P}^-$ with shear rates normalized by the ultrasound-shifted critical shear rate $\dot{\gamma}^*_\mathrm{US}$. 
        (f) Ratio $\dot{\gamma}^*_\mathrm{US}/\dot{\gamma}^*$ as a function of $a_\mathrm{US}$.  
        (g) Illustration of ultrasound de-thickening effect. 
    }
    \label{fig:3scaling}
\end{figure}

To further investigate how large stress fluctuations evolve with increasing ultrasonic amplitude, we performed experiments series at various constant and normalized shear rates $(\dot\gamma-\dot\gamma^{*})/\dot\gamma^*$ ranging from -40 to 30\%, as shown in Fig.~\ref{fig:3scaling}(a). Overall, we observe that the probability of the suspension remaining in the low-stress regime, $\mathcal{P}^-$, increases with ultrasonic amplitude, indicating a more fluid-like behavior.
As shown in Fig.~\ref{fig:3scaling}(b), $\mathcal{P}^-$ can be shifted along the $x-$axis by a factor $a_0$  to collapse on a master curve with respect to a reference taken at $(\dot{\gamma} - \dot{\gamma}^*)/{\dot{\gamma}^*}=0$.
Figure~\ref{fig:3scaling}(c) shows that $a_0$ scales linearly with $(\dot{\gamma} - \dot{\gamma}^*)/{\dot{\gamma}^*}$, indicating that higher $a_\mathrm{US}$ is needed at higher shear rates to maintain the same shear-thickening state.
In addition, Fig.~\ref{fig:3scaling}(d) shows how stress fluctuations evolve with the normalized shear rate $(\dot{\gamma} - \dot{\gamma}^*)/\dot{\gamma}^*$ at fixed $a_\mathrm{US}$. Overall, $\mathcal{P}^-$ decreases with increasing $\dot{\gamma}$, reflecting a greater jamming tendency. To account for the effect of ultrasound, we introduce a critical shear rate $\dot{\gamma}^*_{\mathrm{US}}$ and rescale as $(\dot{\gamma} - \dot{\gamma}^*_{\mathrm{US}})/\dot{\gamma}^*_{\mathrm{US}}$. As shown in Fig.~\ref{fig:3scaling}(e), this scaling collapses the data onto a master curve (reference: $a_\mathrm{US}=0$). Finally, Fig.~\ref{fig:3scaling}(f) shows that $\dot{\gamma}^*_{\mathrm{US}}/\dot{\gamma}^*$ increases monotonically with $a_\mathrm{US}$.
These two perspectives can be unified through the flow curve representation in Fig.~\ref{fig:3scaling}(g). Ultrasound tunes the shear-thickening behavior by increasing the critical shear rate from $\dot\gamma^{*}$ to $\dot\gamma^{*}_\mathrm{US}$, which effectively shifts the flow curve toward higher shear rates. The larger $a_\mathrm{US}$, the more pronounced the shift. Consequently, the high-viscosity region originally associated with DST is displaced toward lower viscosities — achieving an effective de-thickening of the suspension. As a consequence, at higher shear rates, stronger $a_\mathrm{US}$ are required to induce this unjamming effect.

\section{Discussion}


Cornstarch suspensions are well known to exhibit pronounced shear thickening~\cite{Hermes2016,Denn2018}, often interpreted within the Wyart--Cates framework as a stress-driven transition from lubricated to frictional contacts~\cite{wyart2014,Richards2020}. However, recent studies have shown that cornstarch departs significantly from this picture. Even at low shear, particles form adhesive and fragile networks, leading to strong shear-induced stiffening and large temporal fluctuations in viscosity~\cite{gauthier2023}. These adhesive interactions originate from polysaccharide chains coating the granules, which can entangle and form hydrogen-bonded contacts under compression, effectively locking particles into jammed configurations~\cite{brown2014,oyarte2017}. 

At the macroscopic level, shear thickening is characterized by pronounced stress heterogeneities associated with aggregation or density wave~\cite{ovarlez2020,saint2018}. These heterogeneities depend on the confinement ratio relative to particle size. In our geometry, $h/(2r_{\mathrm{cs}})=71$, placing the system in the density-wave regime~\cite{gauthier2023}, where dense regions percolate across the gap and give rise to load-bearing force networks responsible for shear thickening.

Overall, discontinuous shear thickening (DST) in cornstarch suspensions emerges from the formation of adhesive, load-bearing networks embedded within heterogeneous density-wave structures~\cite{gauthier2023,ovarlez2020}. These networks are stabilized by a combination of frictional, normal, and adhesive interactions and persist over timescales much longer than the microscopic forcing times, undergoing intermittent formation and collapse that drive transitions between fluid-like and jammed states~\cite{brown2014,oyarte2017}. A key open question is therefore how an externally imposed high-frequency perturbation, such as ultrasound, can destabilize these long-lived structures and continuously shift the DST transition. We examine below several physical mechanisms that may contribute to this acoustic fluidization.

\vspace{5mm}

\noindent \textbf{Ultrasound vibration versus  particle surface roughness --} 
As suggested by~\cite{sehgal2019}, ultrasound-induced dethickening should occur when the vibration amplitude matches the particle surface roughness, i.e., $a_{\mathrm{US}} \sim \delta h = 2~\mu\mathrm{m}$. This threshold is not sharply observed in our experiments: as shown in Fig.~\ref{fig:3scaling}(f), the critical shear rate $\dot{\gamma}_{\mathrm{US}}^*$ varies continuously as $a_{\mathrm{US}}$ increases from 0 to $4~\mu\mathrm{m}$. However, this does not invalidate the hypothesis, as cornstarch particles are polydisperse, with surface roughness ranging from approximately $0.2$ to $2~\mu\mathrm{m}$. We note, however, that while such criteria may be relevant, it is insufficient on its own, as only gradients of force can destabilize force chains that would otherwise move collectively as a rigid block and remain intact.

 \begin{figure}[htbp]
    \centering
    \includegraphics[width=0.47\textwidth]{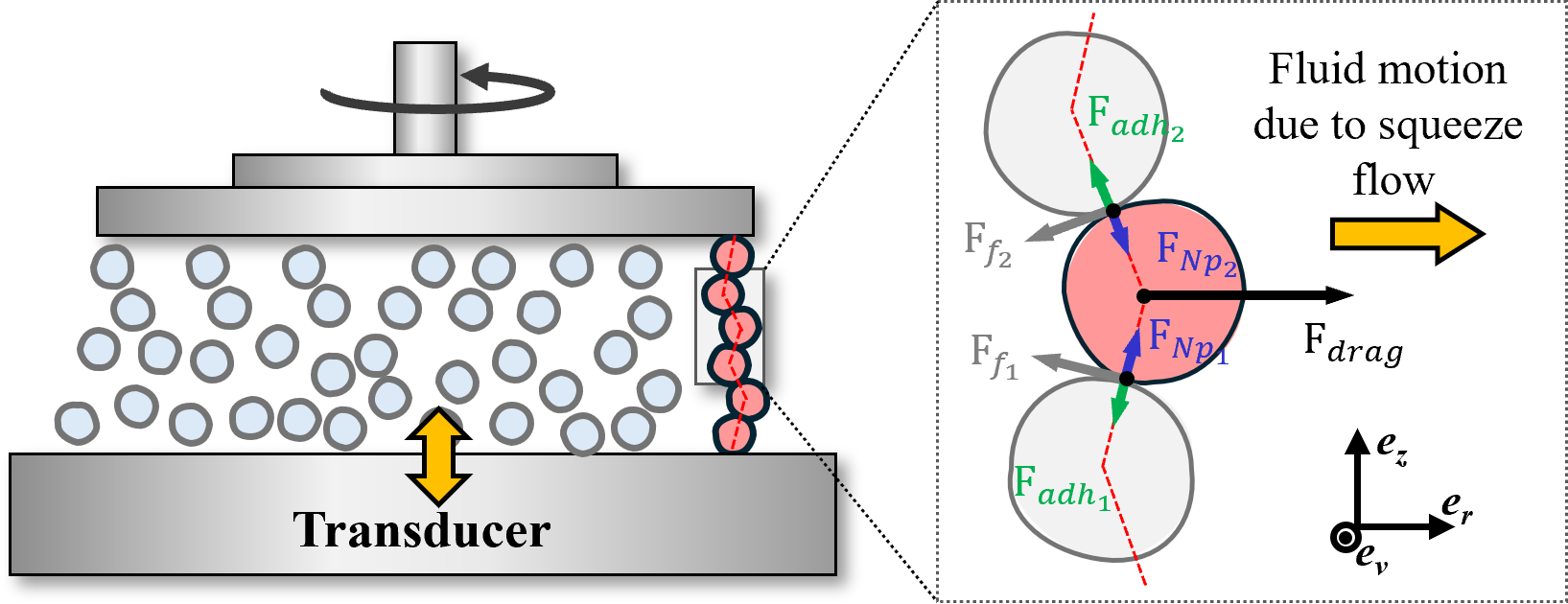}
    \caption{ Schematic diagram of the forces acting on a cornstarch particle embedded in a force chain within a density wave. On short time scales, we assume that the force chain remains immobile while the surrounding fluid undergoes squeeze flow induced by the ultrasound.
    }
    \label{fig:4force}
\end{figure}

\vspace{5mm}
\noindent \textbf{Separation of time scales and porous-medium description --}
We now delve into mechanisms that may destabilize density waves and the associated force chains. As discussed in \cite{dages2021, gibaud2020}, because the ultrasound wavelength ($\lambda\simeq 10$~cm) is much larger than the gap ($h=1$~mm), ultrasound induces oscillatory extensional strain $\gamma^e_{\mathrm{US}} \sim a_{\mathrm{US}}/h$ and radial squeeze flow with strain amplitude $\gamma^s_{\mathrm{US}} \sim Ra_{\mathrm{US}}/(2h^2)$. In our experiments, squeeze flow dominates over extensional flow, with $\gamma^{s}_{\mathrm{US}} / \gamma^{e}_{\mathrm{US}} = R / (2h) \simeq 12$.
\tg{Given the short timescale of the ultrasonic oscillation, $1/f \approx 70~\mu\mathrm{s}$, we hypothesize that the cornstarch force network remains effectively immobile on such timescales, while the interstitial fluid undergoes oscillatory squeeze flow.
This assumption is supported by the observation that, upon a strong reduction of the shear rate from the DST regime to the fluid state, the suspension relaxes over timescales of order $\sim 1$~s (see SM.F~\cite{SM_wang}), in agreement with~\cite{Rijan2017}, and much longer than the ultrasonic period. This indicates a clear separation of time scales between structural relaxation and ultrasonic forcing.
On the ultrasound time scale, we therefore treat the cornstarch dispersion as an effectively immobile porous medium through which the interstitial fluid flows.}
The relative motion between the particle network and the interstitial fluid, directly observed in cornstarch suspensions under shear~\cite{rathee2022}, independently supports this picture. The permeability of the cornstarch matrix is estimated using the Kozeny--Carman equation~\cite{rehman2024}, where $\phi$ is the particle volume fraction: 
$K = \frac{(1 - \phi)^3 (2r_\mathrm{cs})^2}{180\, \phi^2} \simeq 0.99~\mu\mathrm{m}^2.$

\vspace{5mm}

\noindent \textbf{Role of acoustic streaming --}
\tg{Acoustic streaming is a steady fluid flow generated by high-frequency sound waves through nonlinear interactions with the fluid and boundaries~\cite{zarembo1971}. It can structure dispersed systems~\cite{pignon2024,bosson2025} and has been proposed as a possible mechanism for unjamming in shear-thickening suspensions~\cite{barth2026}.} In the present case, however, its contribution is expected to be limited, as the fluid is confined within a porous matrix whose pore size ($a_p \simeq \sqrt{K} \approx 1~\mu$m) is significantly smaller than the viscous penetration depth ($\delta = \sqrt{\eta/(\pi \rho f)} \approx 4.6~\mu$m; see SM.E~\cite{SM_wang})~\cite{muller2013,karlsen2015,happel2012} suppressing the formation of acoustic streaming at the pore scale.
Nevertheless, the heterogeneous nature of shear thickening gives rise to 
transient fluid pockets coexisting with jammed regions~\cite{rathee2022, 
saint2018}, within which localized streaming flows may develop and 
progressively erode adjacent force-chain structures.

\vspace{5mm}

\noindent \textbf{Force-chain destabilization by flow-induced gradients --}
Let us know focus on the high-frequency interstitial squeeze flows generated by the ultrasound wihin the porous matrix, and analyze how these flows couple to the porous structure impact the mechanical response of the system.

A key parameter for characterizing interstitial flow in porous media is the 
\emph{permeability Reynolds number}~\cite{wood2020}, defined as $\mathrm{Re}_K = \rho v_0 \sqrt{K}/\eta$, where $v_0 = \pi f R a_\mathrm{US}/ h$ is the characteristic flow velocity. One obtains $\mathrm{Re}_K >1$ when $a_\mathrm{US}>  1~\mu$m separating linear Darcy from the Forchheimer regime.
The first hypothesis, within the Darcy  regime, proposes that the squeeze flow promotes slippage of the density wave at the rheometer interfaces (rotor and stator), thereby enhancing the overall flow. Following~\cite{lang2021,lang2024}, we computed the time-resolved velocity profile of the squeeze flow generated by one ultrasonic oscillation. As shown in SM.G~\cite{SM_wang}, the velocity profile is close to the one of a plug flow. This generates a strong velocity gradient near the plates on short length scales ($\simeq 10~\mu$m) that may promote the detachment of shear-jammed cornstarch particles from the rheometer boundaries, supporting the hypothesis that slip drives the shear unjamming induced by ultrasonic vibrations.

The second hypothesis explores the Forchheimer regime and posits that the squeeze flow destabilizes the force chains within the bulk of the density wave. In this regime, flow through the porous media becomes spatially heterogeneous: fluid preferentially travels through certain pore channels and experiences local accelerations or decelerations. 
Experimentally, in porous systems visualized via confocal imaging, Datta et al.~\cite{Datta2013} reported that velocity magnitudes in individual pores and their components follow broad distributions and exhibit correlation lengths controlled by pore geometry. 
Typically, velocity fluctuations \cite{hill2002, lu2018,arthur2018,dukhan2014} are of the order of $\Delta v/v_0 \sim 0.1\text{–}0.3.$ and persist over several pore to pore distances.
Let us now estimate the forces acting on a single cornstarch particle within a force chain, as illustrated in Fig.~\ref{fig:4force} and tabulated in SM.H~\cite{SM_wang}. The particle’s stability is maintained by adhesive and local normal and frictional forces, while destabilization arises from drag forces induced by velocity fluctuations of the ultrasound-driven squeeze flow. 
\tg{The adhesion force, measured by atomic force microscopy, is $F_{\mathrm{adh}}\simeq 20$~nN~\cite{oyarte2017}. The normal force per particle, considering a total rotor force $F_N\simeq 0.25$~N and a mean field approach (see SM.H~\cite{SM_wang}), is $F_N^{(p)}\simeq 20$~nN. The corresponding frictional force, using a frictional coefficient $\mu=0.5$, is $F_f\simeq 10$~nN. }

At the onset of the Forchheimer regime, for $a_\mathrm{US} = 1~\mu\mathrm{m}$, when inertial and viscous drag forces are of comparable magnitude, a velocity 
fluctuation of $\Delta v = 0.12\,v_0$ already produces a gradient in viscous force $\Delta F_\mathrm{drag} = 6\pi\eta r_\mathrm{cs}\Delta v$, comparable to the stabilization force scale per particles. \tg{Recent simulations have shown that force chains in cornstarch-like suspensions with rolling constraints are predominantly linear structures~\cite{sharma2026}, making them particularly sensitive to force gradients along their backbone: a spatially inhomogeneous drag force creates torque imbalances that a linear chain cannot redistribute laterally, unlike a clustered or branched network}. We therefore estimate that ultrasound-induced drag-force fluctuations are sufficient to create force and torque imbalances along these linear chains on ultrashort timescales, destabilizing them and promoting flow over jamming.

\medskip
\noindent\tg{\textbf{Shift of the DST threshold --}
Beyond the disruption of individual contacts, our results suggest that ultrasound alters the stability of force networks. The onset of DST is governed by a balance between stabilizing interactions (frictional, normal, and adhesive forces) and destabilizing perturbations. Ultrasound-induced interstitial flows introduce additional fluctuating drag forces, which act as a source of mechanical noise at the particle scale. As a result, larger stresses or shear rates are required to stabilize load-bearing structures, leading to a continuous shift of the DST transition. In this framework, acoustic fluidization can be interpreted as a renormalization of the force balance governing shear thickening.}

\section{Conclusion}
\tg{We demonstrate experimentally that the apparent viscosity reduction reported in the literature under high-power ultrasound~\cite{sehgal2019, sehgal2024, ong2024jamming} is not an independent effect, but instead results from a shift in the onset of DST in dense adhesive suspensions. The collapse of stress probability distributions onto master curves shows that ultrasound progressively drives the system toward more fluid-like states.
While our modeling identifies different dominant mechanisms at different $a_{\mathrm{US}}$, the experiments reveal no sharp threshold. Instead, the continuous evolution of $\dot{\gamma}^*_{\mathrm{US}}$ with $a_{\mathrm{US}}$ (Fig.~\ref{fig:3scaling}) indicates that fluidization arises from multiple concurrent pathways whose relative importance evolves smoothly.
Our interpretation relies on a separation of time scales between ultrasonic oscillations ($\sim 70~\mu$s) and suspension relaxation ($\sim 1$~s), allowing the shear-thickened state to be viewed as an effectively immobile porous matrix. Within this framework, fluidization can originate from boundary slip, bulk force-chain destabilization driven by drag-force fluctuations, and localized acoustic streaming in transient fluid regions.
Beyond the disruption of individual contacts, our results suggest that ultrasound modifies the stability of force networks by introducing fluctuating hydrodynamic forces at the pore scale. As a result, larger stresses or shear rates are required to sustain load-bearing structures, leading to a continuous shift of the shear-thickening transition.
Together with orthogonal oscillatory shear~\cite{Lin2016, Ness2018}, ultrasound provides a complementary route to control shear thickening. While oscillatory shear acts at the macroscopic scale, ultrasound operates through local forcing within the pore space, making it particularly suited to confined geometries such as pipes and extruders.
More broadly, ultrasound offers an in situ strategy to fluidize both attractive gels~\cite{gibaud2020, dages2021} and dense adhesive suspensions. This capability opens new perspectives for controlling flow instabilities in complex materials such as cementitious slurries, where localized unjamming may be achieved without interrupting operation.}

\section*{Author Contributions}
\noindent AW carried out the experiments. AW and TG analyzed and interpreted the data. AW and TG wrote the paper. TG and FT designed/managed the project.
 
\section*{Conflicts of interest}
\noindent There are no conflicts to declare.

\section*{Data Availability}
\noindent The data supporting the findings of this study are available from the authors upon reasonable request

\section*{Acknowledgements}
\noindent The authors are grateful for valuable discussions with Annie Colin, Valeria Garbin, Sébastien Manneville and Di Bao.
This work was supported by the European Union’s Horizon Europe Framework Programme (HORIZON) under the Marie Skłodowska-Curie Grant Agreement 101120301. This work benefited from meetings within the French working group GDR CNRS 2019 "Solliciter LA Matière Molle" (SLAMM).

%


\clearpage
\section*{Supplemental Material}
\renewcommand{\figurename}{figure}
\renewcommand{\thefigure}{S\arabic{figure}}
\setcounter{figure}{0}

Supplemental materials for the article titled "Acoustic modulation of shear thickening transition in dense adhesive suspensions", authored by Aoxuan Wang, Fabrice Toussaint and Thomas Gibaud, we detail the following:
A) Cornstarch volume fraction;
B) Cornstarch particle size and surface roughness;
C) Rheo-Ultrasound apparatus; 
D) Critical shear rate and normal forces;
E) Acoustic streaming;
F) Relaxation of the shear thickening state;
G) Velocity profile of the interstitial fluid in the Darcy regime due to squeeze flow;
H) Summary of the physical quantities.

\subsection{Cornstarch volume fraction}
We work with cornstarch particles (Sigma-Aldrich) dispersed in water and density-matched using cesium chloride, following the protocol in~\cite{han2016,saint2018}.
To convert the mass concentration $c_w$ of cornstarch suspension into volume fraction, we follow the methodology described by Han \cite{han2016}. First, we compute an effective volume fraction $\phi_v$ that accounts for the volume occupied by dry cornstarch particles and the suspending fluid, taking into consideration the intrinsic moisture content of cornstarch grains. This is given by:

\[
\phi_v = \frac{(1 - \beta) \, m_{\mathrm{cs}} / \rho_{\mathrm{cs}}}
{(1 - \beta) \, m_{\mathrm{cs}} / \rho_{\mathrm{cs}} + m_l / \rho_l + \beta \, m_{\mathrm{cs}} / \rho_w}
\]

where $m_{\mathrm{cs}}$ and $m_l$ are the masses of cornstarch and liquid, respectively; $\rho_{\mathrm{cs}} = 1.63\, \mathrm{g/cm^3}$, $\rho_l = 1.63\, \mathrm{g/cm^3}$, and $\rho_w = 1.00\, \mathrm{g/cm^3}$ are the densities of cornstarch, suspending liquid, and water; and $\beta = 0.11$ is the estimated mass fraction of moisture in the dry cornstarch. For a suspension with a cornstarch mass concentration of $c_w=39$\,wt\%, this yields $\phi_v \approx 0.338$.

Since the porous starch granules absorb a portion of the suspending liquid, the actual solid volume fraction $\phi$ is corrected using a prefactor $(1 + \alpha)$, where $\alpha = 0.3$ quantifies the additional fluid uptake. 
The final volume fraction is thus given by $\phi = (1 + \alpha)\,\phi_v \approx 1.3 \times 0.338 \approx 0.439$, indicating that a $c_w=39$~wt\% cornstarch suspension corresponds to a true volume fraction of approximately $\phi =44$~\%.

\subsection{Cornstarch particles}

\begin{figure}[htbp]
    \centering
    \includegraphics[width=.45\textwidth]{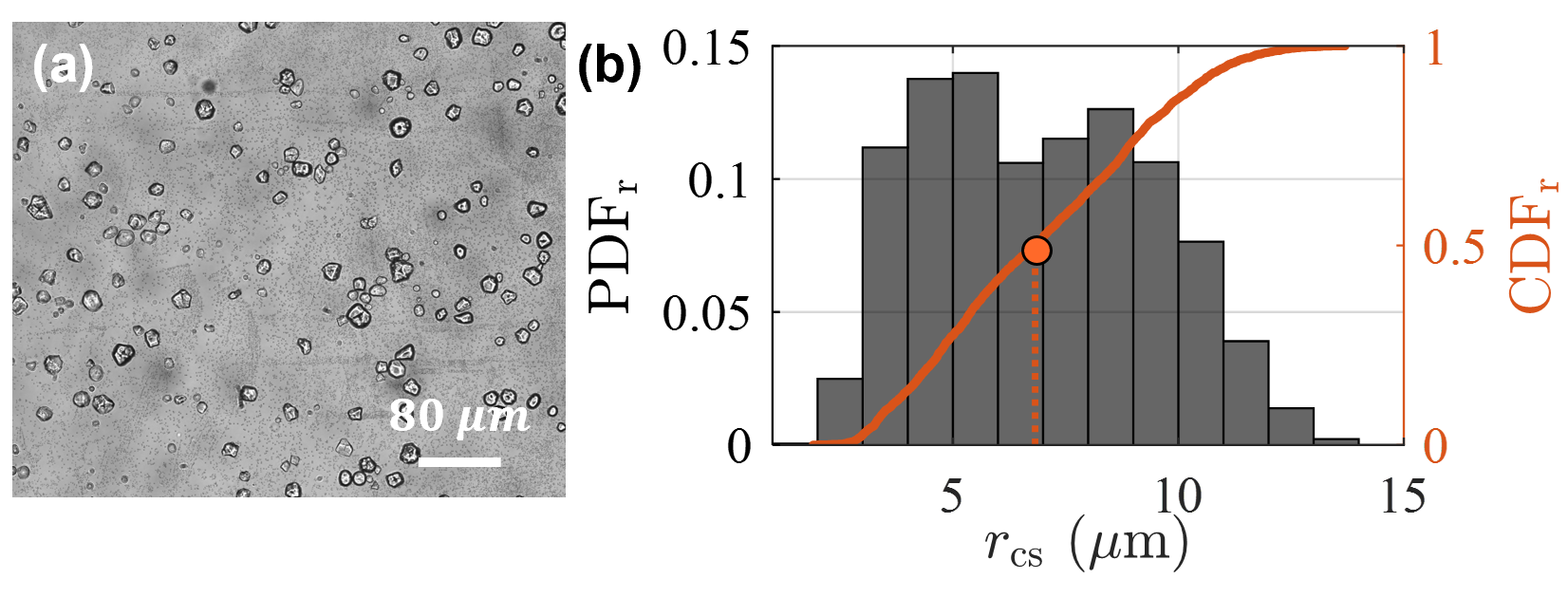}
    \caption{
    Particle size distribution of cornstarch.
    (a) Bright field microscopy image of a dilute suspension of cornstarch. 
    (b) Probability distribution function of the Particle diameter $\mathrm{(PDF_\mathrm{r})}$ obtained from microscopy images (in gray) and cumulative distribution function $\mathrm{(CDF_\mathrm{r})}$ (in orange). Dotted lines indicate the mean particle radius $\langle r_\mathrm{cs}\rangle = 7 $~$\mu$m.
    }
    \label{fig:S1radius}
\end{figure}

\begin{figure}[htbp]
    \centering
    \includegraphics[width=.45\textwidth]{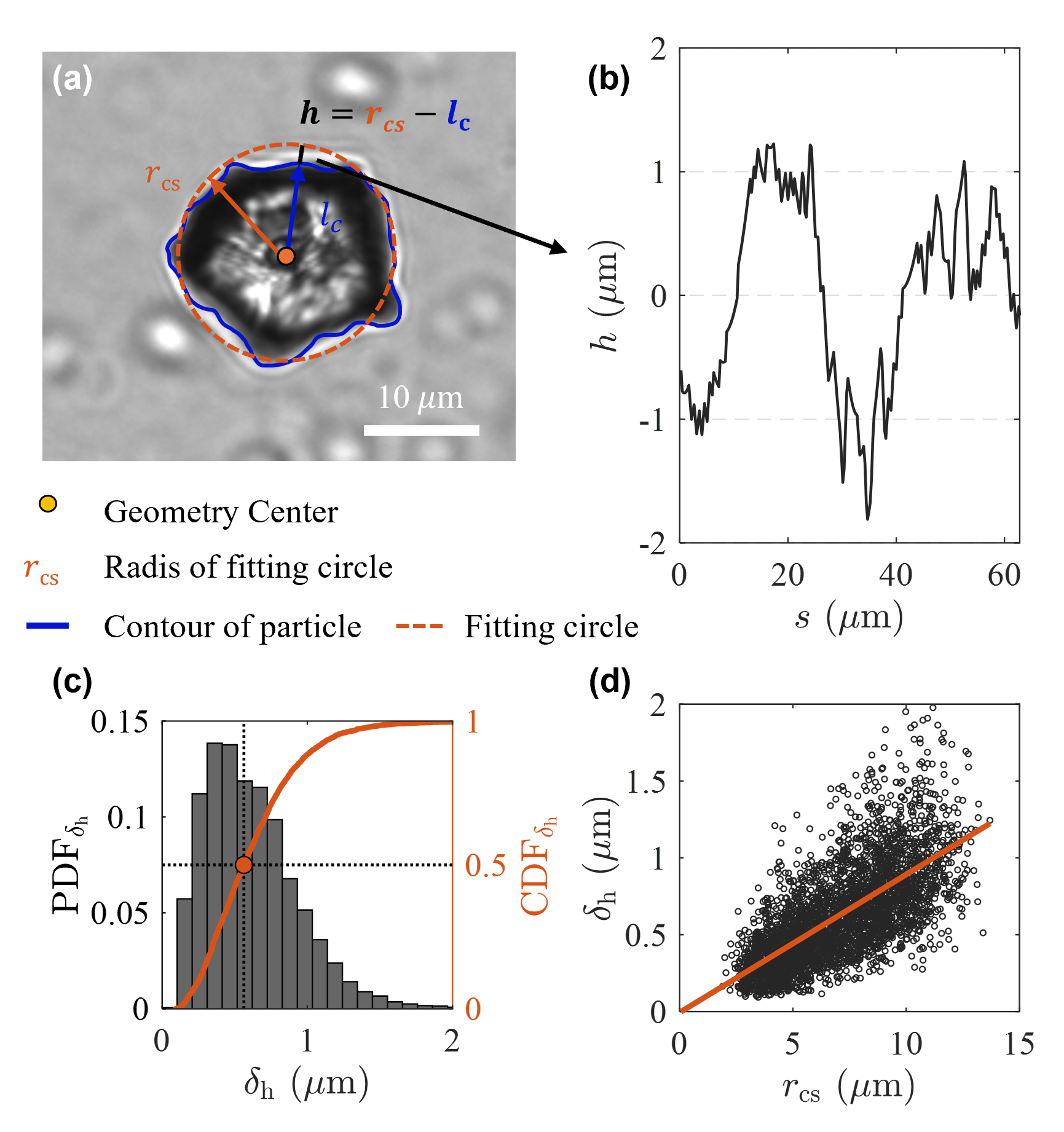}
    \caption{
    Contour roughness of the particles measured by bright field microscopy images.
    (a) Bright field microscopy image of one single corn starch particle. The blue line correspond to the contour length $s$ of the particle and its height with respect to the particle center is $l_c$. The dash red line corresponds to the mean particle circumference: $r_\mathrm{cs}=\langle l_\mathrm{c} \rangle$.
    (b) The height of the roughness $h=r-l_\mathrm{c}$ along the contour length $s$. We note $\delta_\mathrm{h}=std(h)$ the particle roughness. 
    (c) Probability distribution function (gray), $\mathrm{PDF}_{\mathrm{\delta_{h}}}$, and cumulative distribution function (orange) ,$\mathrm{CDF}_{\mathrm{\delta_h}}$ as a function of $\delta_\mathrm{h}$. 
    (d) Particle roughness $\delta_\mathrm{h}$ as function of the mean radius of the particles, $r_\mathrm{cs}$. (c-d) are statistical measured and obtained over 3900 particles.
    }
    \label{fig:S2roughness}
\end{figure}

Using bright field microscopy on dilute dispersion of cornstarch suspensions dispersed in water, we could image isolate single cornstarch particles that sedimented at the bottom of the observation cell as shown in~Fig~\ref{fig:S1radius}(a) and \ref{fig:S2roughness}(a).
We then developed a MATLAB algorithm  to quantitatively analyze the geometric features of each particles. Specifically, the program extracts particle contours profiles $s$ from binarized images and determines their geometric centers. Each contour is approximated as a circle, with the fitted radius $r_\mathrm{cs}$ defined as the mean distance of contour points from the geometric center $l_\mathrm{c}$. The radial deviations of contour points from this fitted circle $h$ are then calculated, and their standard deviation $\delta_\mathrm{h}$ is used to characterize the contour roughness of each particle. The program further performs statistical analyses of particle radii and contour roughness, generates probability and cumulative distribution functions, and examines the correlation between these two parameters.

In the particle's local frame, as illustrated in Fig.~\ref{fig:S2roughness}(a), we define $l_c$ as the radial distance from the center to the contour. The average value $\langle l_c \rangle$ gives the mean particle radius $r_\mathrm{cs}$. The probability density function of $r$, denoted $\mathrm{PDF_r}$, confirms that the particles are polydisperse, with a mean radius of $r_\mathrm{cs} = 10~\mu\mathrm{m}$.

In Fig.~\ref{fig:S2roughness}(b), we define the local roughness height along the particle contour as $h = r - l_\mathrm{c}$. The surface roughness $\delta_\mathrm{h}$ is then defined as the standard deviation of $h$: $\delta_\mathrm{h} = \mathrm{std}(h)$. The probability density function $PDF_{\delta_h}$ shown in Fig.~\ref{fig:S2roughness}(c) indicates that the roughness is polydisperse, with an average value of $\delta_\mathrm{h} = 0.56~\mu\mathrm{m}$. Interestingly, as shown in Fig.~\ref{fig:S2roughness}(d), we find that the roughness scales linearly with the particle radius $r_\mathrm{cs}$ , following $\delta_h = 0.09\, r_\mathrm{cs}$.

\begin{figure*}[htbp]
    \centering
    \includegraphics[width=.91\textwidth]{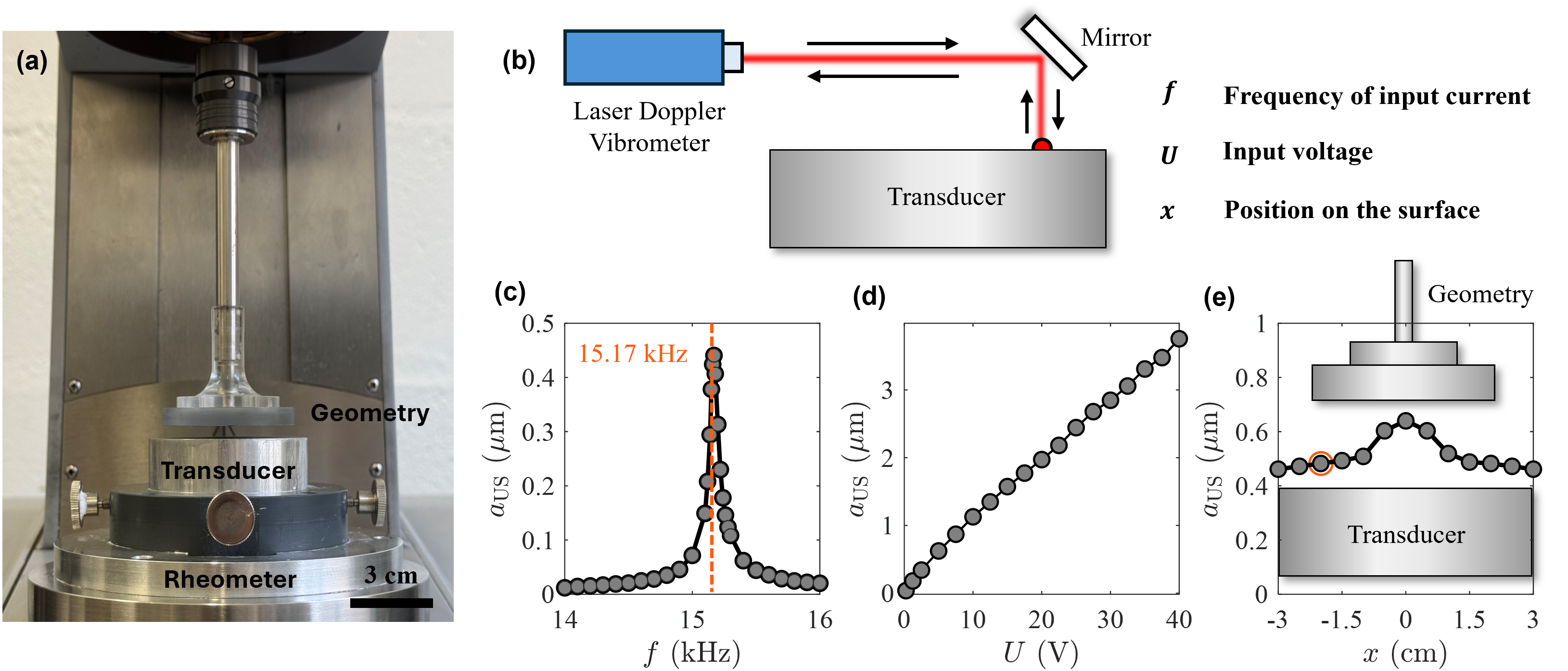}
    \caption{
    Rheo-Ultrasound apparatus and its calibration.
    (a) Photography of the Rheo-Ultrasound setup.
    (b) We used a vibrometer to determine the vibration amplitude of the transducer  $a_\mathrm{US}$ as a function of frequency $f$, position with respect to its center $x$ and input voltage $U$.
    (c) Amplitude of transducer $a_\mathrm{US}$ as a function of frequency $f$. The dash line indicates the resonant frequency at 15.17~kHz.
    (d) Amplitude of transducer $a_\mathrm{US}$ as a function of input voltage $U$.
    (e) Amplitude of transducer $a_\mathrm{US}$ as a function of the position along the transducer surface $x$.
    (c) and (d) were measured at 1~cm away from the edge of the transducer as the orange circle shown in (e).
    }

    \label{fig:S3calib}
\end{figure*}

\subsection{Rheo-Ultrasound apparatus}
The Rheo-Ultrasound apparatus (Fig.~\ref{fig:S3calib}(a)) consists of an upper part—a stress-controlled Anton Paar rheometer equipped with a plexiglass plate of radius 25~mm—and a lower part, the stator, which comprises a piezoelectric transducer (ThorLab PKT40B) of radius 30~mm. The transducer is driven by an amplified sinusoidal voltage $U$ (amplifier: ES-HSA42014), converting electrical input into vertical surface vibrations of amplitude $a_{\mathrm{US}}$. The gap between the plates is fixed at $h=1$~mm.

We calibrate the transducer using a laser Doppler vibrometer (Polytec OVF-505), measuring (i) its resonance frequency $f$ (Fig.~\ref{fig:S3calib}(c)), (ii) the amplitude of ultrasonic vibration $a_{\mathrm{US}}$ as a function of input voltage $U$ (Fig.~\ref{fig:S3calib}(d)), and (iii) the spatial profile of $a_{\mathrm{US}}$ along the radial direction of the transducer surface (Fig.~\ref{fig:S3calib}(e)).

In Fig.~\ref{fig:S4temp}(a), we show the time evolution of the temperature measured at various ultrasonic amplitudes $a_{\mathrm{US}}$. Temperature is recorded using a sensor (National Instruments, NI USB-TC01) affixed to the transducer, while the gap is filled with light mineral oil to ensure thermal contact. 
\begin{figure}
    \centering
    \includegraphics[width=.43\textwidth]{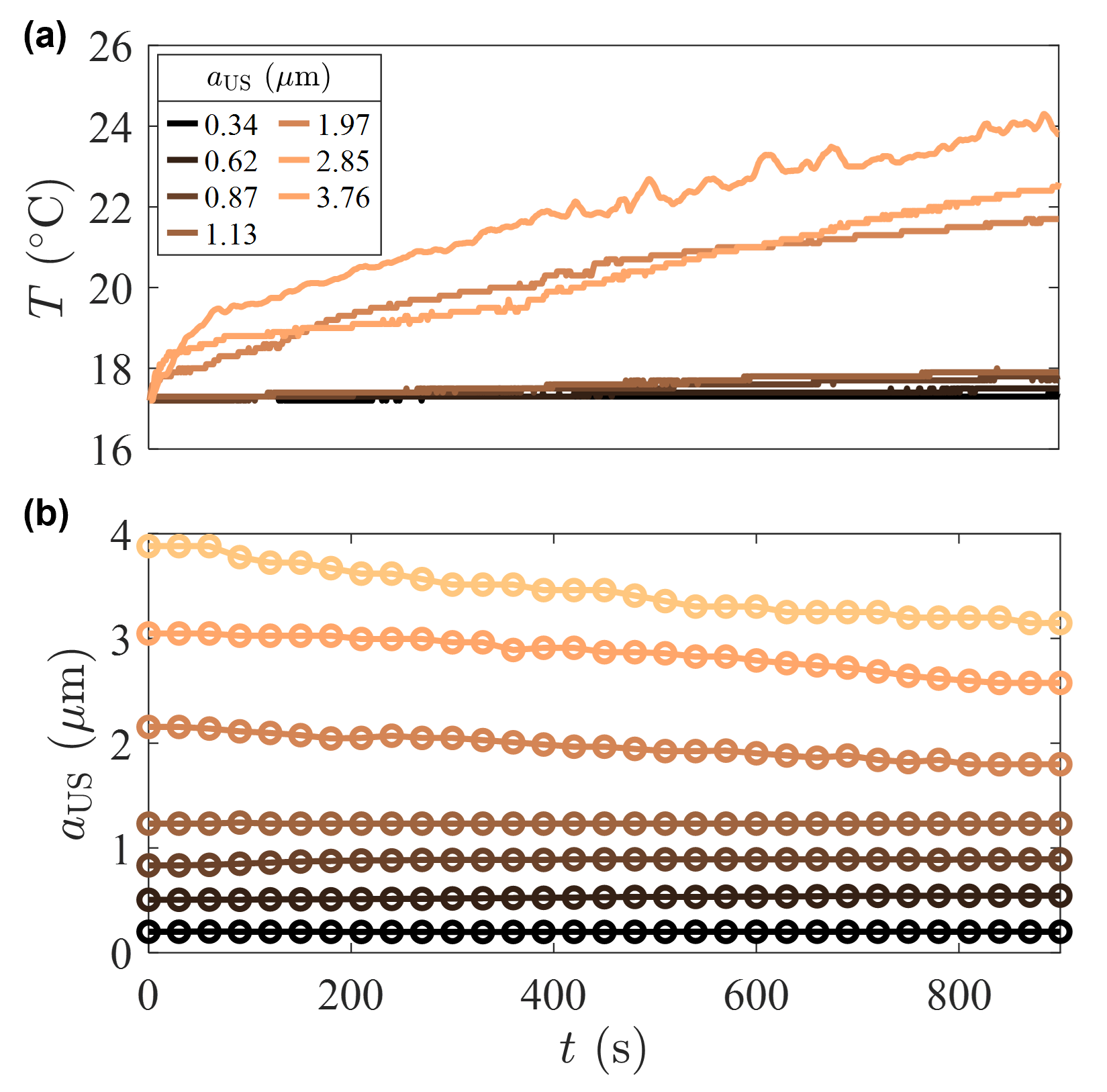}
    \caption{
    Temperature calibration.
    (a) Time evolution of the temperature $T$ of the sample subject to ultrasound of amplitude $a_\mathrm{US}$. 
    (b) Time evolution of  $a_\mathrm{US}$.
    }
    \label{fig:S4temp}
\end{figure}
In Fig.~\ref{fig:S4temp}(b), we track the time evolution of $a_{\mathrm{US}}$ during operation. The experiments described in this study last for 150~s—a duration over which both temperature and $a_{\mathrm{US}}$ remain nearly constant, indicating thermal and mechanical stability during measurements.

\subsection{Critical shear rate and normal forces }

\begin{figure}
    \centering
    \includegraphics[width=.47\textwidth]{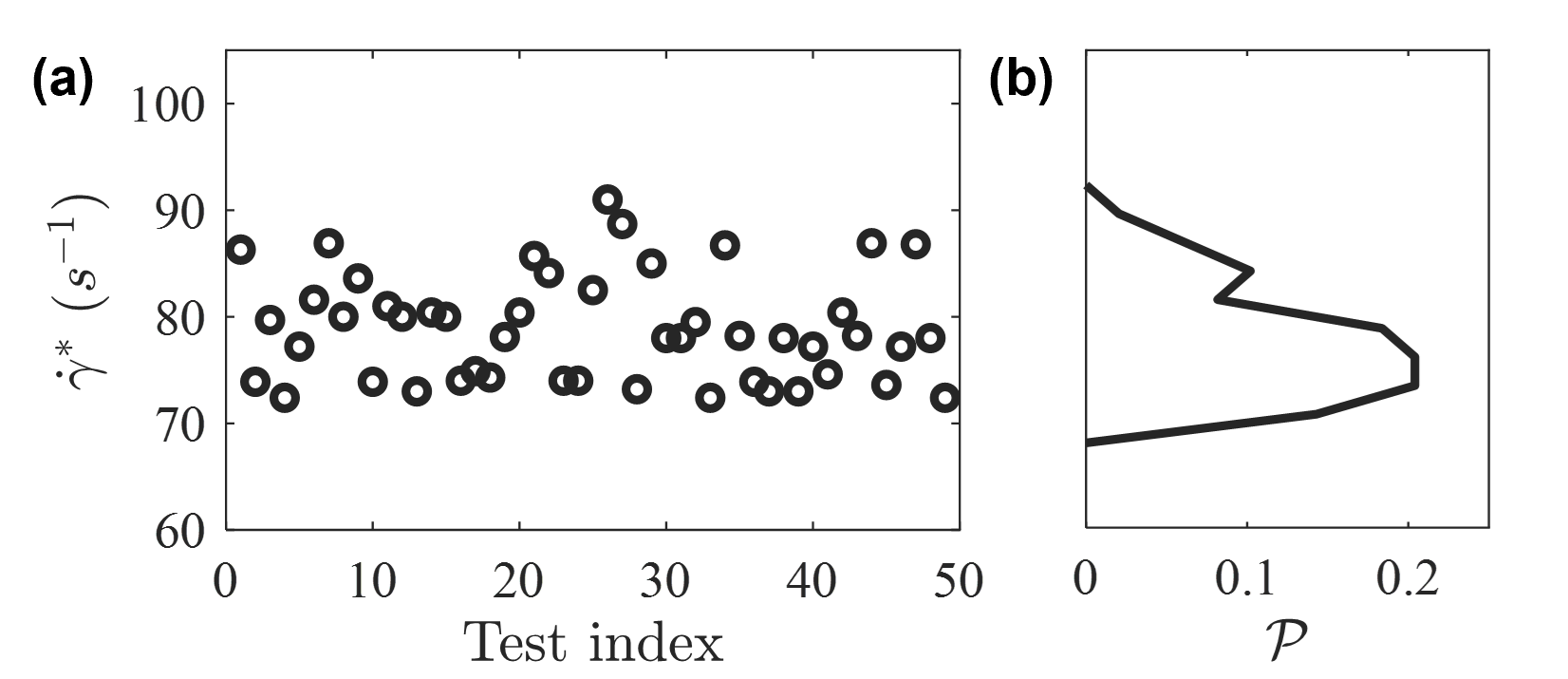}
    \caption{critical shear rate 
    (a) Critical shear rate $\dot{\gamma^*}$ as function of the experiment number.
    (b) Probability distribution function of $\dot{\gamma^*}$ 
    }
    \label{fig:S5sr}
\end{figure}

Figure~\ref{fig:S5sr} shows the critical shear rate $\dot{\gamma}^*$ extracted during the flow curve step of the rheological protocol (Fig.~1(c) and Fig.~2(a)). We define $\dot{\gamma}^*$ as the shear rate corresponding to the threshold stress $\sigma_t = 28$~Pa.

\begin{figure*}
    \centering
    \includegraphics[width=.9\textwidth]{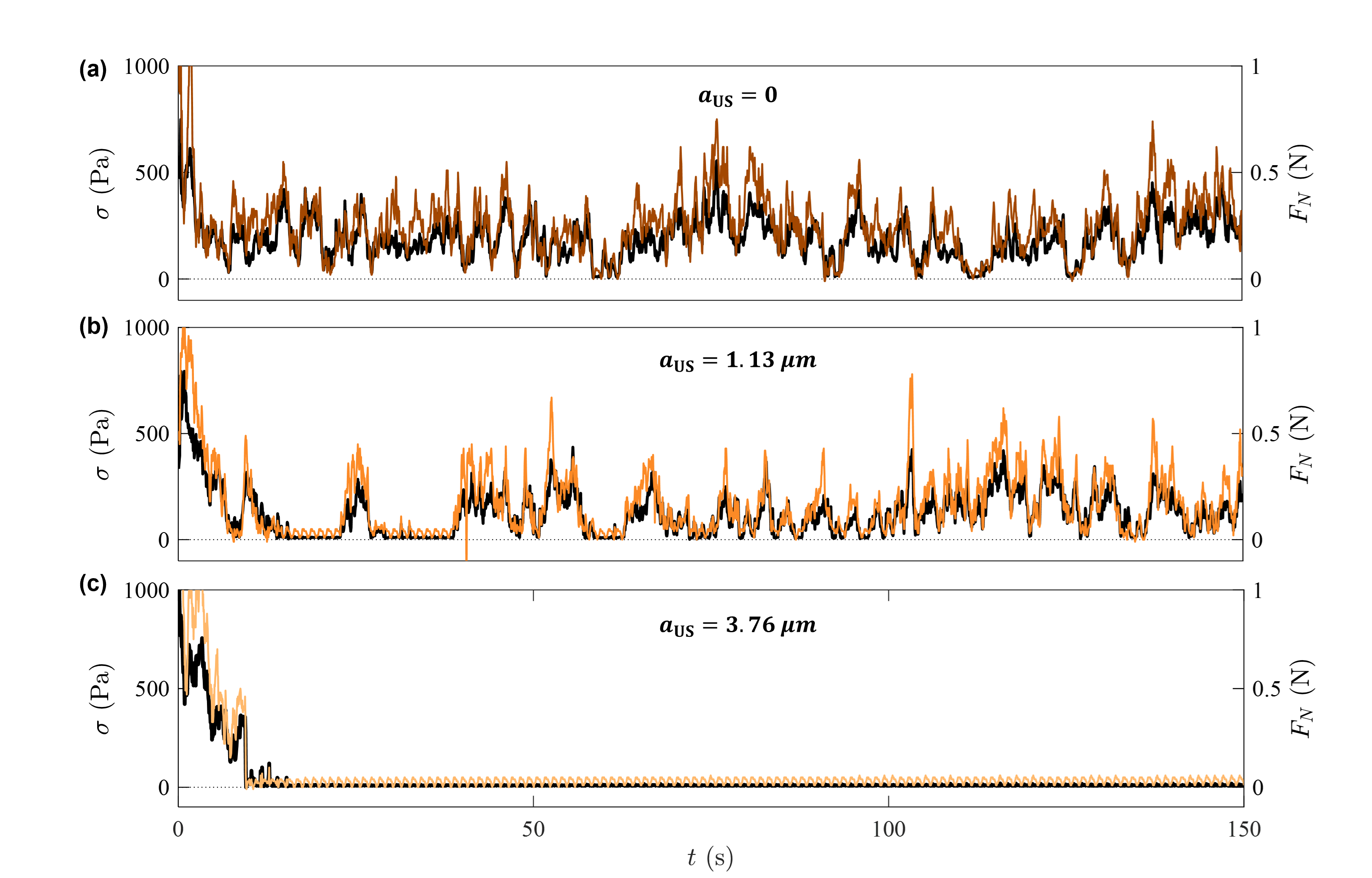}
    \caption{
    Normal force and shear stress as functions of time at different ultrasound amplitude: (a) $a_\mathrm{US}$=0, 
    (b) $a_\mathrm{US}$=1.13~$\mu$m, and (c) $a_\mathrm{US}$=3.76~$\mu$m.
    The black lines represent the shear stress $\sigma$, and the lines with color represent normal force. $F_N$
    }
    \label{fig:S6NormalForce}
\end{figure*}

Figure~\ref{fig:S6NormalForce} displays the time evolution of the stress $\sigma$ and the normal force $F_N$. Ultrasound of amplitude $a_{US}$=0, 1.13 and 3.76~$\mu$m are turned on at $t=10$~s.

\vspace{3 mm}
\subsection{Acoustic streaming}
Acoustic streaming is a net, steady fluid flow generated by high-frequency sound waves through nonlinear interactions with the fluid and boundaries~\cite{zarembo1971}. The viscous penetration depth is the distance over which a fluid may develop acoustic streaming: $\delta=\sqrt{\eta/(\pi\rho f)}=4.6$~$\mu$m. $\delta$ being larger than the pore size ($a_p\simeq\sqrt{K}=0.99~\mu$m), the ultrasound-induced fluid motion barely penetrates the pores and we therefore neglect acoustic streaming as direct source to destabilize density waves and the chain forces.

\noindent\textbf{Acoustic streaming forces.} Let us derive a simple estimation of the streaming force density based on \cite{muller2013,karlsen2015}. Acoustic streaming originates from the time–averaged nonlinear advection term in the
Navier--Stokes equation.  Writing the fluid velocity as the sum of a fast oscillatory
acoustic component $\mathbf{v}$ and a slow steady streaming component $\mathbf{U}$,
\[
\mathbf{u} = \mathbf{v} + \mathbf{U}, \qquad |\mathbf{U}| \ll |\mathbf{v}|,
\]
and averaging over one acoustic period, the driving force per unit volume for the steady
flow is given by the Reynolds stress
\[
\mathbf{f}_{\mathrm{s}} = -\Big\langle \rho\,(\mathbf{v}\cdot\nabla)\mathbf{v} \Big\rangle .
\]
In the vicinity of rigid boundaries, the oscillatory velocity varies over the viscous
penetration depth $\delta = \sqrt{2\eta/(\rho\omega)}$, so that
$\nabla \mathbf{v} \sim v_a/\delta$, where $v_a$ is the acoustic particle velocity
amplitude.  This yields the scaling estimate $
\mathbf{f}_{\mathrm{s}}
\;\sim\;
\rho\,\frac{v_a^2}{\delta}.
$

The steady streaming flow $\mathbf{U}$ produced by this forcing is opposed by viscous
friction, whose magnitude is $\eta\nabla^2\mathbf{U}\sim \eta U_s/\delta^2$.
In steady state, the nonlinear driving must balance this viscous resistance,
\[
\eta\,\frac{U_s}{\delta^2}
\;\sim\;
\rho\,\frac{v_a^2}{\delta},
\]
implying $U_s \sim (\rho v_a^2 \delta)/\eta$, the classical Rayleigh streaming velocity.
The corresponding streaming force density is then obtained from the viscous term,
$f_{\mathrm{s}} \sim \eta U_s/\delta^2$.  Substituting $U_s$ gives $
f_{\mathrm{s}} \sim \frac{\rho v_a^2}{\delta}.
$
To express this in terms of the acoustic pressure amplitude $p_0$, we use the linear
acoustic relation $v_a = p_0/(\rho c)$, where $c$ is the sound speed.
This finally yields the estimate
$$
f_{\mathrm{s}} \sim \frac{\eta\, v_a^{\,2}}{c\,\delta^{2}}
$$
which highlights that acoustic streaming in viscous fluids is controlled by
(i) the acoustic particle velocity $v_a$, (supposing perfectly radiating transducer  in water we substitute $v_a$ by transducter wall velocity $v_a \approx v_{\rm wall} = 2\pi f\,a_{\mathrm{US}}$);
(ii) the boundary-layer thickness $\delta$; (iii) the sound speed $c$,
and (iv) the viscosity $\eta$, reflecting the fundamentally viscous origin of the
streaming force. At $a_{US}=4~\mu$m, on a volume defined by $V_{\delta}=\pi R^2\delta$ the streaming force is $F_s=V_{\delta} f_s\simeq4\times 10^4$~nN.

\vspace{3 mm}
\noindent\textbf{Effect of confinement.}
Let us now estimate how this streaming force is damped when confined to a pore~\cite{happel2012}.
We consider steady streaming driven by the time-averaged Reynolds stress of an oscillatory acoustic field.
The nonlinear driving (source) term per unit volume scales as
\[
f_{\rm drive}\sim \rho\frac{v_a^2}{\delta},
\]
where $v_a$ is the acoustic particle velocity amplitude and $\delta$ is the viscous penetration depth 
(so that $\nabla v\sim v_a/\delta$). In free boundary-layer streaming the steady streaming velocity $U_{\rm free}$ 
is set by a balance between this driving and viscous resistance acting over the boundary-layer thickness,
$\eta\frac{U_{\rm free}}{\delta^2}\sim f_{\rm drive}\sim \rho\frac{v_a^2}{\delta},
$
hence the classical Rayleigh scaling
$$
U_{\rm free}\sim\frac{\rho v_a^2\delta}{\eta}.
$$

Now consider a cylindrical pore of characteristic radius $a\ll\delta$.  Inside the pores, the steady streaming flow
is constrained by the pore cross section, so the dominant viscous resistance scales with the pore length scale,
i.e. $\nabla^2 U\sim U/a^2$. Balancing the same driving with this pore-scale viscous resistance gives
$
\eta\frac{U_{\rm pore}}{a^2}\sim f_{\rm drive}\sim \rho\frac{v_a^2}{\delta},
$
and therefore
$$
U_{\rm pore}\sim \frac{\rho v_a^2 a^2}{\eta\,\delta}.
$$

Taking the ratio of the pore velocity to the free (unconfined) streaming velocity yields the damping factor
$
\frac{U_{\rm pore}}{U_{\rm free}}
\sim
[\dfrac{\rho v_a^2 a^2}{\eta\,\delta}]/[\dfrac{\rho v_a^2 \delta}{\eta}]
 = 
\frac{a^2}{\delta^2}.
$
Since streaming force densities scale as viscous resistance $\sim \eta U/\ell^2$ with the same characteristic length,
the streaming force inside the pore is reduced by the same factor,
$
\frac{f_{\rm pore}}{f_{\rm free}}\sim\frac{a^2}{\delta^2}.
$

In our case, the streaming force is damped by a factor $(a_p/\delta)^2=0.05$. At the pore scale (Volume, $V_p=4/3\pi a_p^3$), the damped force is therefore $F_s=f_s(a_p/\delta)^2 V_p \simeq3\times 10^{-4}$~nN a value much smaller than all the other forces considered to destabilize the cornstarch chain forces even at the highest ultrasonic amplitude.

Nevertheless, the heterogeneous nature of shear thickening gives rise to 
transient fluid pockets coexisting with jammed regions~\cite{rathee2022, 
saint2018}, within which localized streaming flows may develop and 
progressively erode adjacent force-chain structures.

\subsection{Relaxation of the shear thickening state}
Comparing the time scales of ultrasound and rheometer-induced shear is crucial for understanding how ultrasound affects the mechanism of shear thickening. Fig.~\ref{fig:S7relax} shows the relaxation time of the corn starch suspension as it recovers from the shear-thickened state, which is approximately 1~$s$ in agreement with~\cite{Rijan2017}, much longer than the ultrasound period of 66~$\mu$s. This indicates that the force chains responsible for shear thickening can be considered quasi-static on the time scale of the ultrasonic excitation.

\begin{figure}
    \centering
    \includegraphics[width=.45\textwidth]{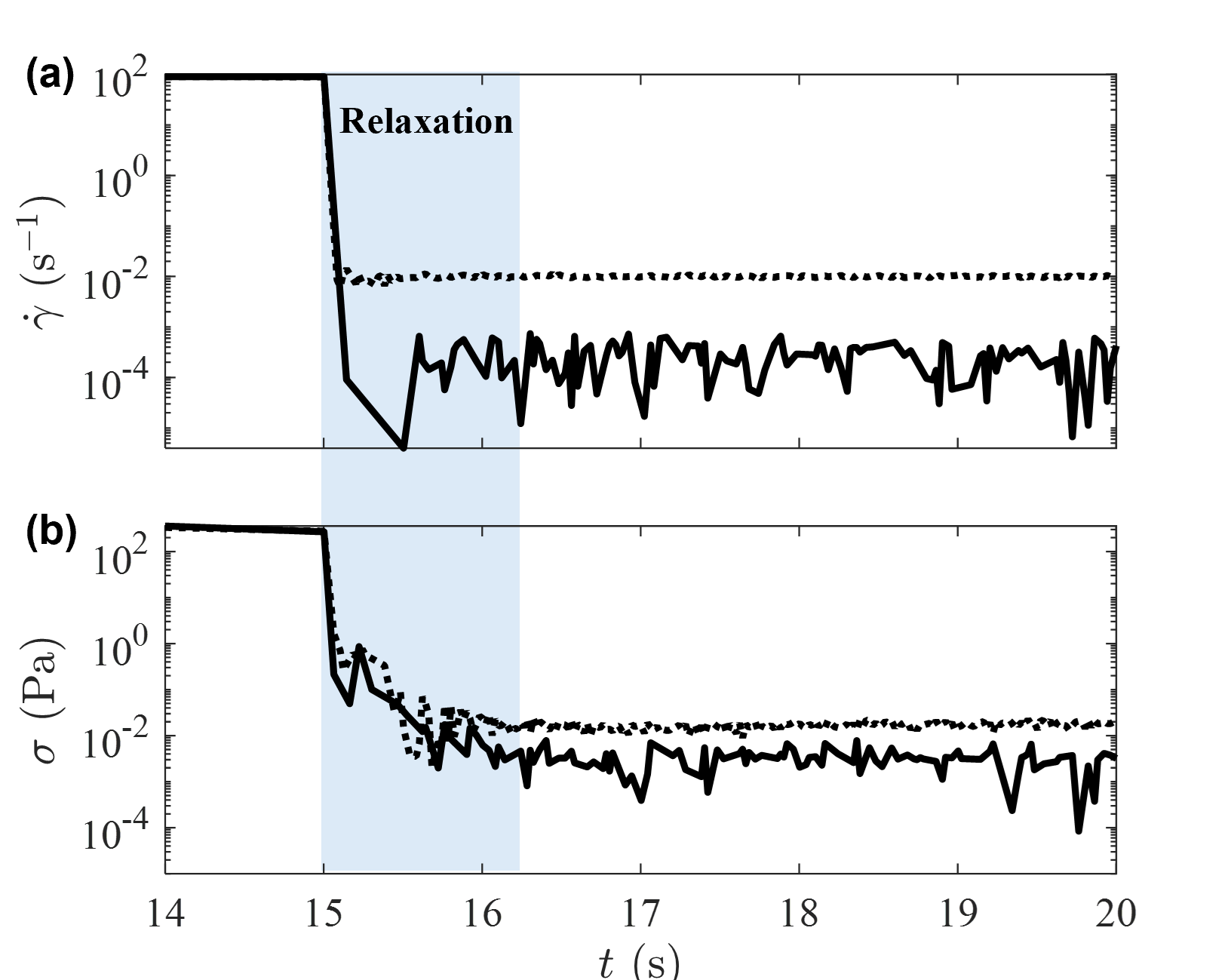}
    \caption{
    Relaxation time of corn starch dispersion from a shear-thickening state to flow state. We perform a similar protocol with previous experiments Fig. \ref{fig:1fc}(d).(a) Shear rate is set to 100 ~s$^{-1}$ for 15~s and then dropped to 0~s$^{-1}$ (dash line) or 0.01~s$^{-1}$ (line). The data acquisition frequency is 50~Hz. (b) stress response. 
    }
    \label{fig:S7relax}
\end{figure}

\subsection{Velocity profile of the interstitial fluid in the Darcy regime due to squeeze flow}
\begin{figure*}[htbp]
    \centering
    \includegraphics[width=.9\textwidth]{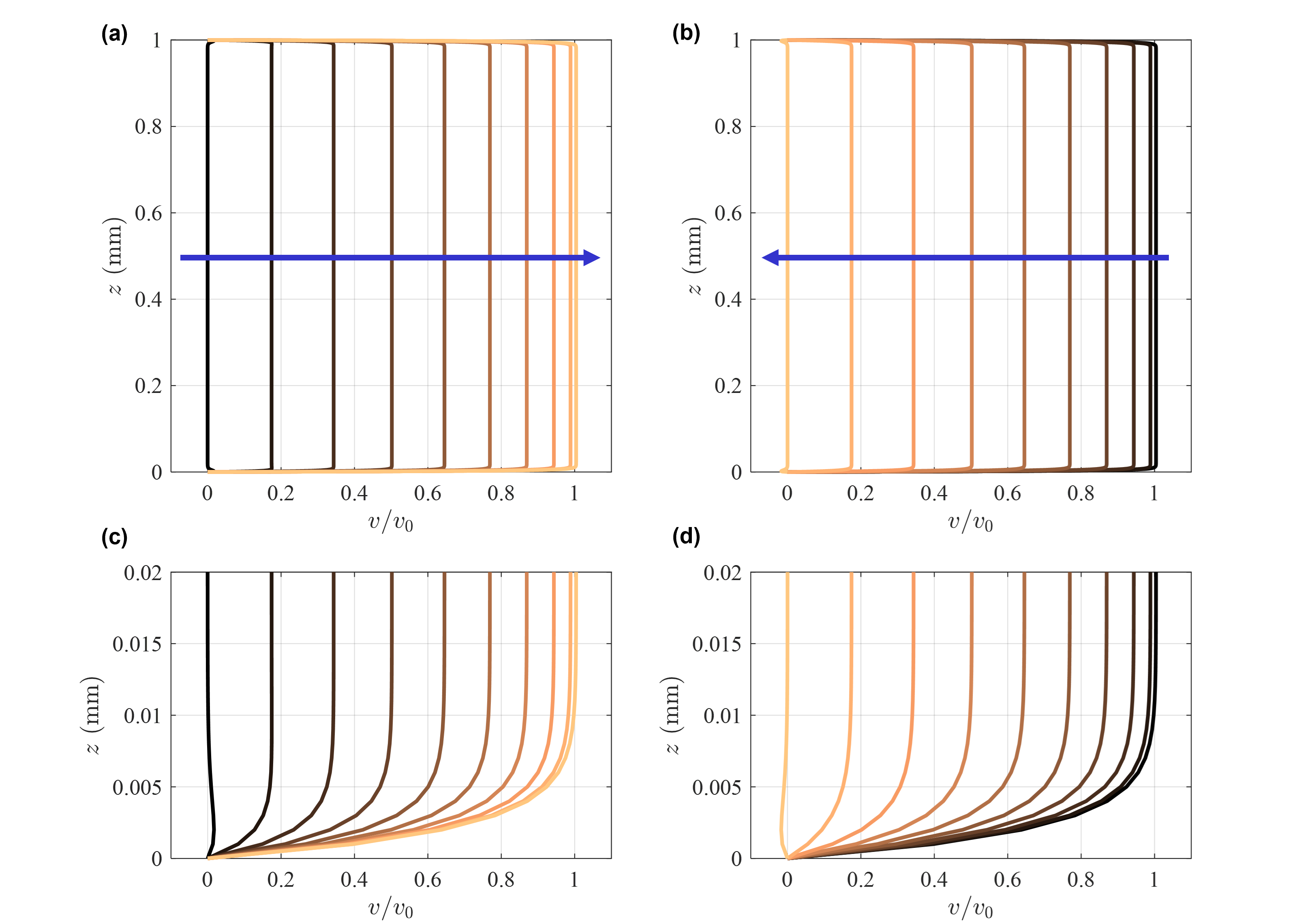}
    \caption{
 Velocity profiles of the interstitial fluid along the height direction $z$ of the gap at the edge of the geometry ($r = R$) generated by high-frequency oscillatory ultrasound (US) of amplitude $a_\mathrm{US}$. The time evolution of $v(z,t)/v_0$ along $z$ is shown during (a) the first quarter cycle and (b) the second quarter cycle of a squeeze flow oscillation. Panels (c) and (d) show a zoomed view of $v(z,t)/v_0$ near $z=0$, where large velocity gradients occur. The direction of the blue arrow indicates the progression of time.
    }
    \label{fig:S8velocity}
\end{figure*}
The radial squeeze flow generated in a dense cornstarch suspension by vertical harmonic excitation is intrinsically complex. On short time scales, we model the motion as a fluid within a porous matrix subjected to a squeeze flow. The small gap height renders viscous effects dominant and the flow laminar, while corn starch matrix modeled by a porous media of permeability $K$ introduces Darcy drag and the transient driving makes inertial corrections relevant. The Darcy–Lapwood–Brinkman (DLB) framework ~\cite{lang2021,lang2024} captures this competition and yields the time evolution of the oscillatory velocity field profile of the interstitial fluid as a function of the gap height $z$ in the thin-gap limit.
We assume that: (i) the flow through the porous medium is laminar, so that Darcy’s law applies (negligible pore-scale inertia and small Reynolds number, $Re<1$); (ii) the fluid is incompressible and Newtonian, with constant viscosity $\eta$, density $\rho$, and kinematic viscosity $\nu$; (iii) the porous medium is homogeneous and isotropic, with constant permeability $K$; and (iv) there is no slip at the boundaries

In this case, the velocity is primarily governed by two dimensionless numbers: $\mathrm{Wo} = \sqrt{\frac{h^2}{t_0 \nu}} \simeq 154$, which quantifies the ratio of inertial to viscous effects, and $\mathrm{Br} = \sqrt{\frac{h^2}{K}} \simeq 502$, which quantifies the ratio of Darcy drag to viscous diffusion. Within the framework, the calculated velocity is displayed in Fig.~\ref{fig:S8velocity}. We observe that the velocity displays a plugflow-like profile within the bulk of the gap and a strong gradient of velocity near the boundaries at $z=0$. This strong velocity gradient could promote the detachment of the cornstarch matrix from the transducer or the rotor.

\subsection{Summary of the physical quantities}
The tab.~\ref{tab:forces} recapitulated the physical quantities used in the paper.

\begin{table*}[t]
\centering
\renewcommand{\arraystretch}{1.3}
\begin{tabular}{p{10cm} | p{3.5cm} | p{2.5cm}}
\hline
\multicolumn{3}{l}{\textbf{a) Rheometer}} \\
\hline
Geometry: parallel plates, radius  & $R$ & $25~\mathrm{mm}$  \\
and gap &  $h$ &  $1~\mathrm{mm}$   \\
Control variable: shear rate & $\dot{\gamma}$ &   \\
Measured : shear stress, normal force & $\sigma,\; F_N$ &  \\
\hline

\multicolumn{3}{l}{\textbf{b) Ultrasound}} \\
\hline
Resonance frequency & $f$ & $15.17~\mathrm{kHz}$ \\
Wavelength in water (c=1500~m/s) & $\lambda = c/f$ & $10~\mathrm{cm}$ \\
Amplitude (from calibration)& $a_{\mathrm{US}}$ & $0	\rightarrow4\,\mu\mathrm{m}$ \\
Transducer surface velocity & $v_{\mathrm{wall}} = 2\pi f\, a_{\mathrm{US}}$ & $0 	\rightarrow 0.38~\mathrm{m/s}$ \\
Acoustic power & $P = 2\pi^{2} \rho c (\pi R^2) f^2 a_{\mathrm{US}}^2$ & $0 	\rightarrow 214$~W \\
Oscillatory extensional strain & $\gamma^e_{\mathrm{US}} = a_{\mathrm{US}}/h$ & $0 	\rightarrow 4\times 10^{-3}$ \\
Radial squeeze-flow strain & $\gamma^s_{\mathrm{US}} = R a_{\mathrm{US}}/(2 h^2)$ & $0 	\rightarrow 0.05$ \\
Strain ratio & $\gamma^s_{\mathrm{US}} / \gamma^e_{\mathrm{US}} \sim R/(2h)$ & $12.5$ \\
\hline

\multicolumn{3}{l}{\textbf{c) Cornstarch}} \\
\hline
Particle radius & $r_{\mathrm{cs}}$ & $7~\mu\mathrm{m}$ \\
Contour roughness & $\delta_h$ & $0.1	- 0.5~\mu$m \\
Weight Concentration & $c_w$ & $39\%$ \\
Volume fraction & $\phi$ & $44\%$ \\
Adhesion force & $F_{\mathrm{adh}}$ & $20~\mathrm{nN}$~\cite{oyarte2017} \\
Typical normal force on the rotor in the DST regime & $F_N$ & $0.25~\mathrm{N}$ \\
Normal force per particle (see "Note on the normal force per particle" for the justification) & $F_{Np} = \left(\frac{r_{\mathrm{cs}}}{R}\right)^2$ & $20~\mathrm{nN}$ \\
Friction force (with $\mu=0.5$) & $F_f = \mu F_{Np}$ & $10~\mathrm{nN}$  \\
Solvent density & $\rho$ & $10^3~\mathrm{kg/m^3}$ \\
Solvent viscosity & $\eta$ & $1~\mathrm{mPa\cdot s}$ \\
Solvent kinematic viscosity & ${\nu=\eta/\rho}$ & $10^{-6}~\mathrm{m^2/ s}$ \\
\hline

\multicolumn{3}{l}{\textbf{d) Cornstarch as a porous medium}} \\
\hline
Permeability (Kozeny–Carman) with $\phi=0.44$
& $K = \frac{(1-\phi)^3 (2r_{\mathrm{cs}})^2}{180\phi^2}$ 
& $0.99~\mu\mathrm{m}^2$ \\
Pore size & $a_p = \sqrt{K}$ & $1~\mu\mathrm{m}$ \\
\hline

\multicolumn{3}{l}{\textbf{e) Acoustic streaming}} \\ 
\hline
Viscous penetration depth & $\delta = \sqrt{\frac{\eta}{\pi \rho f}}$ & $4.6~\mu\mathrm{m}$ \\
Streaming force density, $v_{\mathrm{wall}}\simeq v_a$ & $f_{\mathrm{s}} = \frac{\eta  v_a^{ 2}}{c \delta^{2}}$ &   \\
Unconfined streaming force with $ V_{\delta}=\pi R^2\delta$ & $F_s^{\mathrm{unconf}} = f_sV_{\delta}$ & $0 	\rightarrow 4\times 10^{4}$~nN \\
Confined streaming force with $V_p$ the volume of a pore& $F_s=f_s(a_p/\delta)^2 V_p$ &  $0 	\rightarrow 3\times 10^{-4}$~nN \\
\hline

\multicolumn{3}{l}{\textbf{f) Squeeze flow in a porous media}} \\
\hline
Ratio of inertial to viscous effects  & $\mathrm{Wo} = \sqrt{\frac{h^2}{t_0 \nu}}$  & 154\\
Ratio of Darcy
drag to viscous diffusion & $\mathrm{Br} = \sqrt{\frac{h^2}{K}}$  & 502\\

Characteristic velocity  & $v_0 = \frac{\pi f R a_{\mathrm{US}}}{h}$ & $0 	\rightarrow 4.8~\mathrm{m/s}$ \\
Permeability Reynolds number & $\mathrm{Re}_K = \frac{\rho v_0 \sqrt{K}}{\nu}$ & $0 	\rightarrow 4.7$ \\
Drag-force fluctuations with $\Delta v=0.1 v_0$  & $\Delta F_{\mathrm{drag}} = 6\pi \eta r_{\mathrm{cs}} \Delta v$ & $0 	\rightarrow 60~\mathrm{nN}$ \\
\hline

\end{tabular}
\caption{Summary of physical quantities, symbols and values/estimation used in the analysis.}
\label{tab:forces}
\end{table*}

\tg{\noindent\textbf{Note on the normal force per particle $F_{Np}$.} 
To estimate the normal force at the particle scale, we introduce a characteristic force per particle $F_{Np}$ by distributing the macroscopic normal force $F_N$ over the rheometer area. This yields $
F_{Np} \sim F_N \left(\frac{r_{\mathrm{cs}}}{R}\right)^2.
$
Using typical experimental parameters ($F_N \simeq 0.25$~N, $r_{\mathrm{cs}} \simeq 7~\mu$m, $R \simeq 25$~mm), we obtain $F_{Np} \sim 20$~nN.
This estimate assumes a homogeneous stress distribution across the sample and that each particle occupies an area of order $\pi r_{\mathrm{cs}}^2$. We emphasize that it therefore provides a lower bound for the force experienced by individual particles. In shear-thickening suspensions, stress transmission is highly heterogeneous and localized within force chains and dense regions, so that only a fraction of particles carry a significant portion of the load. The actual normal forces within force chains are thus expected to exceed this estimate.
Despite these limitations, this scaling provides a useful reference to compare with other force contributions. In particular, this value is comparable to the adhesive force between particles, $F_{\mathrm{adh}} \simeq 20$~nN~\cite{oyarte2017}, indicating that both normal and adhesive interactions contribute to the stability of contacts. Consequently, the relevant condition for destabilization is that fluctuations in the drag force become comparable to this characteristic force scale, $\Delta F_{\mathrm{drag}} \sim  F_{\mathrm{adh}}\sim F_{Np} \sim 20~\mathrm{nN}$.}

\end{document}